\documentclass[iop]{emulateapj}

\shorttitle{Oxygen in the Solar Neighborhood}
\shortauthors{Rodr{\'i}guez \& Delgado-Inglada}

\begin{document}

\title{The Oxygen Abundance in the Solar Neighborhood}

\author{M\'onica Rodr{\'i}guez and Gloria Delgado-Inglada}
\affil{Instituto Nacional de Astrof{\'i}sica, \'Optica y Electr\'onica (INAOE),
Apdo Postal 51 y 216, 72000 Puebla, Mexico}
\email{mrodri@inaoep.mx, gloria@inaoep.mx}

\begin{abstract}
We present a homogeneous analysis of the oxygen abundance in five \ion{H}{2}
regions and eight planetary nebulae (PNe) located at distances lower than 2 kpc
and with available spectra of high quality. We find that both the collisionally
excited lines and recombination lines imply that the PNe are overabundant in
oxygen by about 0.2~dex. An explanation that reconciles the oxygen abundances
derived with collisionally excited lines for \ion{H}{2} regions and PNe with the
values found for B-stars, the Sun, and the diffuse ISM requires the
presence in \ion{H}{2} regions of an organic refractory dust component that is
not present in PNe. This dust component has already been invoked to explain the
depletion of oxygen in molecular clouds and in the diffuse interstellar
medium.
\end{abstract}


\section{Introduction}

The ionized gas that surrounds young stars in \ion{H}{2} regions and evolved
stars in planetary nebulae (PNe) is subjected to very similar processes, but the
gas in \ion{H}{2} regions samples the present interstellar medium (ISM), while
the gas in PNe samples the ISM of several gigayears ago, when the progenitor
stars formed. Hence, if we choose an element whose abundance in the PN has not
changed significantly from the original one, like oxygen at near-solar
metallicities (e.g., \citealt{kar10}), we can calculate its abundance in
\ion{H}{2} regions and PNe using the same procedure, and compare the differences
with those predicted by galactic chemical evolution.

This potential is somewhat marred by the existing discrepancy between the
abundances derived using collisionally excited lines (CELs) and those implied by
recombination lines (RLs) of the same elements
\citep{pei93}. In all the ionized nebulae studied so far, RLs imply
abundances higher than those implied by CELs by factors around 2 in most cases,
with some PNe showing much higher discrepancies \citep{liu06}. The emissivities
of CELs have a stronger dependence on temperature than those of RLs, and
most of the proposed explanations of the difference rely on the production of
temperature fluctuations by some mechanism in the observed nebulae. The
exception would be those explanations that consider uncertainties in the
recombination coefficients of the heavy elements. The different explanations of
the discrepancy imply that the real abundances will be either close to the
values implied by CELs, closer to the abundances derived with RLs, or
intermediate between them \citep[and references therein]{tor03,rod10}.
Therefore, a meaningful comparison between the abundances derived for \ion{H}{2}
regions and PNe must not only take into account the results provided by both
CELs and RLs, but also the fact that the explanation can be different for the
two kinds of objects.

A further issue to consider is that the ionizing radiation fields can be very
different in \ion{H}{2} regions and PNe. This implies that the corrections for
the contribution of unobserved ions to the total abundance, generally based on
the relative abundances of two ions of another element, are likely to
introduce a different bias in each kind of object. Hence, the best option is to
perform the comparison using an element whose dominant ionization states are all
observed. In \ion{H}{2} regions and low-ionization PNe, this happens with
oxygen. Besides, the required [\ion{O}{2}] and [\ion{O}{3}] lines can be
observed in the optical region of the spectrum along with the other \ion{H}{1}
and diagnostic lines needed for the analysis. This reduces the uncertainties
introduced when comparing the relative intensities of lines measured at widely
separated wavelengths, with mismatched apertures, and with different telescopes.

In this letter, we present a comparative analysis of the oxygen abundance in
five \ion{H}{2} regions and eight low-ionization PNe of the solar neighborhood.
The solar neighborhood is defined here as a region around the Sun with a radius
of 2 kpc. This allows us to secure a representative number of objects whose
abundances should not be much affected by the Galactic abundance gradient. The
analysis follows the same method and uses the same atomic data for all the
objects. We use the best available spectra and provide results for both CELs and
RLs. The derived oxygen abundances are compared with those implied by nearby
young stars and those based on absorption lines in the diffuse ISM.

\section{The sample}
\label{sample}

The sample objects were selected from the compilation in \citet{del09} of
Galactic \ion{H}{2} regions and low ionization PNe with available deep optical
spectra. There are around 100--800 detected lines in each object and the
spectral resolution in the blue is better than 1.5 \AA. All the objects have
individual distance determinations locating them at distances below 2 kpc (see
Section~\ref{an} below). The distance to NGC~6884 could be larger, but at a
Galactic longitude of $82\deg$, the effect of the Galactic radial abundance
gradient should be small.

The PNe have been classified as type II of Peimbert \citep{pei78}, though
NGC~6210 could be of type III \citep{qui07}. \citeauthor{qui07} estimate that
the thin disk progenitors of type II PNe have ages around 4--6 Gyr and initial
masses of 1.2--2.4 M$_\odot$.

\section{Analysis and Results}
\label{an}

We used the same set of lines for the analysis of all the sample objects. The
lines were measured with the same telescope and aperture for each object. The
physical conditions and ionic abundances were calculated with the {\sl nebular}
package in IRAF\footnote{IRAF is distributed by NOAO, which is operated by AURA,
Inc., under cooperative agreement with NSF.}. In order to check the effect of
the atomic data on the calculations, we performed two sets of
calculations. The first set is based on the atomic data compiled in IRAF; the
second one on the atomic data used in the photoionization code Cloudy
(version 08.00, last described by \citealt{fer98}). Below we present the results
of the second set and comment on the differences found with the first one.

The adopted electron densities, $n_\mathrm{e}$, are weighted averages of the
values implied by the diagnostics [\ion{S}{2}] $\lambda6716/\lambda6731$,
[\ion{Cl}{3}] $\lambda5517/\lambda5537$, and [\ion{Ar}{4}]
$\lambda4711/\lambda4740$. For M17 and the lower ionization objects we do not
use the last diagnostic. In M17 the intensity of [\ion{Ar}{4}] $\lambda4740$ is
very uncertain and the intensity ratio $\lambda4711/\lambda4740$ is out of
bounds; in M16 and M20 [\ion{Ar}{4}] $\lambda4740$ was not measured;  in IC~418,
the [\ion{Ar}{4}] line ratio implies $n_\mathrm{e}=4800$~cm$^{-3}$, in
disagreement with the other diagnostics, based on lines whose intensities in
this object are larger by factors $\ge50$. We derive two values for the electron
temperature, $T_\mathrm{e}$, one for the high-ionization regions in the nebulae,
$T_\mathrm{e}$([\ion{O}{3}]), based on the ratio of line intensities
$(\lambda4959+\lambda5007)/\lambda4363$; another for the low-ionization regions,
$T_\mathrm{e}$([\ion{N}{2}]), based on $(\lambda6548+\lambda6583)/\lambda5755$.
We list in Table~\ref{tab1} the physical conditions derived for each object. We
also give the observational uncertainties, i.e., those derived from the
propagation of errors in the line intensities.

\begin{deluxetable*}{lllllllc}
\tabletypesize{\footnotesize}
\tablecolumns{8}
\tablecaption{Physical Conditions\label{tab1}}
\tablewidth{0pt}
\tablehead{
\colhead{Object} & \colhead{$n_\mathrm{e}$([\ion{S}{2}])}
& \colhead{$n_\mathrm{e}$([\ion{Cl}{3}])} &
\colhead{$n_\mathrm{e}$([\ion{Ar}{4}])} & \colhead{$n_\mathrm{e}$(adopted)} &
\colhead{$T_\mathrm{e}$([\ion{N}{2}])} & \colhead{$T_\mathrm{e}$([\ion{O}{3}])}& 
\colhead{ref}\\
   & \colhead{(cm$^{-3}$)} & \colhead{(cm$^{-3}$)} & \colhead{(cm$^{-3}$)} &
\colhead{(cm$^{-3}$)} & \colhead{(K)} & \colhead{(K)} & }
\startdata
\cutinhead{\ion{H}{2} regions}
\objectname{M8}  & \phn$1500\pm200$ & \phn$2000^{+400}_{-300}$ & \phn$2000^{+6700}_{-2000}$ & \phn$1600\pm200$ & \phn$8500\pm100$ & \phn$8000\pm100$ & 1 \\
\objectname{M16} & \phn$1300^{+300}_{-200}$ & \phn$1300^{+700}_{-600}$ & \phn\nodata & \phn$1300\pm200$ & \phn$8500^{+100}_{-200}$ & \phn$7600^{+100}_{-200}$ & 2 \\
\objectname{M17} & \phn\phn$500^{+200}_{-100}$ & \phn\phn$200^{+400}_{-200}$ & \phn\nodata & \phn\phn$400\pm100$ & \phn$9200^{+200}_{-300}$ & \phn$7900\pm100$ & 1 \\
\objectname{M20} & \phn\phn$300\pm100$ & \phn\phn$300^{+500}_{-300}$ & \phn\nodata & \phn\phn$300\pm100$ & \phn$8500^{+100}_{-200}$ & \phn$7700^{+300}_{-200}$ & 2 \\
\objectname{M42} & \phn$5400^{+3700}_{-1700}$ & \phn$8000^{+700}_{-600}$ & \phn$4900^{+1100}_{-900}$ & \phn$7000\pm500$ & $10100^{+200}_{-300}$ & \phn$8250\pm40$ & 3 \\
\cutinhead{Planetary Nebulae}
\objectname{IC 418}   & $16400^{+}_{-10600}$ & $13400^{+5600}_{-3500}$ & \phn\nodata & $13900\pm4200$ & \phn$8900^{+700}_{-600}$ & \phn$8700^{+400}_{-200}$ & 4 \\ 
\objectname{NGC 3132} & \phn\phn$500^{+200}_{-100}$ & \phn\phn$800^{+500}_{-400}$ & \phn\phn$300^{+600}_{-300}$ & \phn\phn$500\pm100$ & \phn$9700^{+300}_{-200}$ & \phn$9400^{+200}_{-100}$ & 5 \\ 
\objectname{NGC 3242} & \phn$2100^{+600}_{-400}$ & \phn$1300^{+600}_{-500}$ & \phn$2100^{+800}_{-600}$ & \phn$1800\pm300$ & $12400^{+1800}_{-1100}$ & $11600^{+300}_{-200}$ & 5 \\ 
\objectname{NGC 6210} & \phn$3700^{+1400}_{-900}$ & \phn$4100^{+900}_{-800}$ & \phn$6200^{+1100}_{-1000}$ & \phn$4600\pm600$ & $11200^{+400}_{-300}$ & \phn$9500\pm200$ & 6 \\ 
\objectname{NGC 6543} & \phn$5800^{+7400}_{-2300}$ & \phn$6400^{+5200}_{-2700}$ & \phn$3100^{+1800}_{-1300}$ & \phn$3700\pm1400$ & $10300\pm500$ & \phn$7800\pm200$ & 7 \\ 
\objectname{NGC 6572} & $16000^{+33000}_{-6800}$ & $20100^{+4000}_{-2900}$ & $15100^{+2200}_{-1900}$ & $16400\pm1800$ & $12000^{+600}_{-500}$ & $10200^{+300}_{-200}$ & 6 \\ 
\objectname{NGC 6720} & \phn\phn$500^{+200}_{-100}$ & \phn\phn$500^{+500}_{-400}$ & \phn\phn$700^{+600}_{-500}$ & \phn\phn$500\pm100$ & $10600\pm300$ & $10400^{+300}_{-200}$ & 6 \\ 
\objectname{NGC 6884} & \phn$6700^{+4100}_{-2000}$ & \phn$6800^{+1400}_{-1000}$ & \phn$9700^{+1600}_{-1300}$ & \phn$7900\pm900$ & $11600^{+400}_{-300}$ & $10900^{+200}_{-300}$ & 6
\enddata
\tablerefs{{Line intensities from:} (1) \citet{gar07}, (2) \citet{gar06},
(3) \citet{est04}, (4) \citet{sha03}, (5) \citet{tsa03}, (6) \citet{liu04a},
(7) \citet{wes04}.}
\end{deluxetable*}

The $\mbox{O}^{+}/\mbox{H}^+$ abundance ratio was calculated using the values
found for $n_\mathrm{e}$, $T_\mathrm{e}$[\ion{N}{2}] and
$I(\mbox{[\ion{O}{2}]}~\lambda3727)/I(\mbox{H}\beta)$; for the
$\mbox{O}^{++}/\mbox{H}^+$ ratio we used $n_\mathrm{e}$,
$T_\mathrm{e}$[\ion{O}{3}] and
$I(\mbox{[\ion{O}{3}]}~\lambda\lambda4959,5007)/I(\mbox{H}\beta)$. The total
oxygen abundance, (O/H)$_\mathrm{CELs}$, was derived by adding up the O$^{+}$
and O$^{++}$ abundances, and using ionization correction factors (ICFs) based on
the He$^{++}$/He$^{+}$ abundance ratio that provide estimates for the
contribution of ions of higher degree of ionization \citep{kin94}.
Table~\ref{tab2} lists the Galactic coordinates and distances, $d$, to the
objects, the ionic and total oxygen abundances, and the values used for the ICF
(i.e., O/(O$^{+}$+O$^{++}$), see \citealt{del09}).

Table~\ref{tab2} also lists the results implied by RLs. The O$^{++}$ abundances
were derived using the \ion{O}{2} RLs of multiplet 1 (the only one measured in
all the objects) and the recombination coefficients of \citet{sto94}. The
multiplet intensity was calculated correcting for the unobserved lines, when
necessary, using the prescriptions given by \citet{pei05}. A comparison with the
results obtained in the original references, that use all reliable multiplets,
shows small differences in most cases, with a maximum difference of 0.14~dex. We
estimated the total oxygen abundance implied by RLs, (O/H)$_\mathrm{RLs}$, by
assuming the same ionization fractions found with CELs.

\begin{deluxetable*}{llllcllclcc}
\tabletypesize{\footnotesize}
\tablecolumns{11}
\tablecaption{Galactic Coordinates, Distances, and Chemical Abundances:
$\{X^{+i}\}=12+\log(X^{+i}/\mathrm{H}^{+}$),
$\{X\}=12+\log(X/\mathrm{H}$)\label{tab2}}
\tablewidth{0pt}
\tablehead{
\colhead{Object} & \colhead{$l(\degr)$} & \colhead{$b(\degr)$} &
\colhead{$d$(kpc)} & \colhead{ref} & \colhead{\{O$^+\}_\mathrm{CELs}$} &
\colhead{\{O$^{++}\}_\mathrm{CELs}$} & \colhead{ICF}
& \colhead{\{O\}$_\mathrm{CELs}$} & \colhead{\{O$^{++}\}_\mathrm{RLs}$} &
\colhead{\{O\}$_\mathrm{RLs}$}}
\startdata
\cutinhead{\ion{H}{2} regions}
\objectname{M8}  & \phn\phn6 & \phn$-1$ & 1.322 & 1 & $8.30\pm0.04$ & $7.90\pm0.02$ & $1.00$ & $8.45\pm0.03$ & $8.24$ & $8.79$ \\
\objectname{M16} & \phn17 & \phn$+1$ & 1.719 & 1 & $8.41^{+0.05}_{-0.03}$ & $7.93^{+0.05}_{-0.03}$ & $1.00$ & $8.53^{+0.04}_{-0.02}$ & $8.31$ & $8.91$  \\
\objectname{M17} & \phn15 & \phn$-1$ & 1.814 & 1 & $7.70^{+0.07}_{-0.05}$ & $8.47\pm0.03$ & $1.00$ & $8.54\pm0.02$ & $8.73$ & $8.82$  \\
\objectname{M20} & \phn\phn7 & \phn\phn$0$ & 0.816 & 1 & $8.40^{+0.05}_{-0.03}$ & $7.76^{+0.05}_{-0.07}$ & $1.00$ & $8.49^{+0.04}_{-0.03}$ & $8.08$ & $8.82$ \\
\objectname{M42} & 209 & $-19$ & 0.399 & 1 & $7.77^{+0.06}_{-0.04}$ & $8.45\pm0.01$ & $1.00$ & $8.53\pm0.01$ & $8.61$ & $8.68$ \\
\cutinhead{Planetary Nebulae}
\objectname{IC 418}   &    215 & $-24$ & 1.3 & 2 & $8.53^{+0.18}_{-0.19}$ & $8.11^{+0.05}_{-0.08}$ & $1.00$ & $8.67^{+0.14}_{-0.13}$ & $8.21$ & $8.81$ \\ 
\objectname{NGC 3132} &    272 & $+12$ & 0.77 & 3 & $8.38^{+0.05}_{-0.06}$ & $8.55^{+0.02}_{-0.04}$ & $1.02$ & $8.78^{+0.02}_{-0.03}$ & $8.81$ & $9.03$ \\ 
\objectname{NGC 3242} &    261 & $+32$ & 0.55 & 4 & $6.45^{+0.15}_{-0.19}$ & $8.45^{+0.03}_{-0.04}$ & $1.18$ & $8.53^{+0.03}_{-0.04}$ & $8.85$ & $8.93$ \\ 
\objectname{NGC 6210} & \phn43 & $+38$ & 1.57 & 5 & $7.16^{+0.05}_{-0.07}$ & $8.66\pm0.04$ & $1.01$ & $8.68\pm0.04$ & $9.01$ & $9.02$ \\ 
\objectname{NGC 6543} & \phn96 & $+30$ & 1.55 & 4 & $7.10\pm0.11$ & $8.79\pm0.05$ & $1.00$ & $8.80\pm0.05$ & $9.08$ & $9.09$ \\ 
\objectname{NGC 6572} & \phn35 & $+12$ & 1.49 & 5 & $7.38^{+0.07}_{-0.09}$ & $8.61^{+0.04}_{-0.05}$ & $1.00$ & $8.64^{+0.03}_{-0.05}$ & $8.75$ & $8.77$ \\ 
\objectname{NGC 6720} & \phn63 & $+14$ & 0.704 & 6 & $8.18\pm0.05$ & $8.52^{+0.03}_{-0.05}$ & $1.12$ & $8.73^{+0.02}_{-0.04}$ & $8.88$ & $9.09$ \\ 
\objectname{NGC 6884} & \phn82 & \phn$+7$ & 1.56,3.3 & 7,4 & $7.11^{+0.05}_{-0.07}$ & $8.58^{+0.04}_{-0.03}$ & $1.13$ & $8.65\pm0.04$ & $9.00$ & $9.07$ 
\enddata
\tablerefs{(1) \citet{kha05}, (2) \citet{guz09}, (3) \citet{cia99},
(4) \citet{mel04}, (5) \citet{haj95}, (6) \citet{har07}, (7) \citet{pal02}.}
\end{deluxetable*}

The total oxygen abundances derived using the atomic data compiled in IRAF
differ from those shown in Table~\ref{tab2} by a maximum of 0.05~dex with one
exception, IC~418. For this nebula, the value implied by IRAF is $\sim0.2$~dex
higher. This large difference arises because the nebula has a high density and
because its oxygen abundance is dominated by O$^+$. When the electron density is
high, the absolute uncertainties in $T_\mathrm{e}$([\ion{N}{2}]) and
$n_\mathrm{e}$ are large, and since the estimated O$^+$ abundance is very
sensitive to the physical conditions, its value is subject to large
fluctuations.

In fact, as pointed out by the referee, the values of
$T_\mathrm{e}$([\ion{N}{2}]) are usually affected by many uncertainties, and
this could be critical for our purposes. The [\ion{N}{2}] diagnostic ratio is
very susceptible to errors in the flux calibration and reddening correction, and
can be altered by a contribution from recombination to
[\ion{N}{2}]~$\lambda5755$ \citep{rub86}, or by contamination from high-density
objects included in the slit, like cometary knots, globules, proplyds, or
Herbig-Haro objects \citep[e.g.,][]{mes08}. The recombination contribution can
be large for objects with high degrees of ionization, but for these objects
O$^+$ makes a small contribution to the total abundance, which is then barely
affected. On the other hand, the contamination by high-density objects could be
a problem for the sample PNe, since most of them were observed using a long-slit
scanning all of their volume. Our sample \ion{H}{2} regions are less likely to
be contaminated, since they were observed using small slits
($\sim3''\times10''$), and include nebulae like M42, the Orion Nebula, where
proplyds and Herbig-Haro objects are easily resolved and identified.
The uncertainties in $T_\mathrm{e}$([\ion{N}{2}]) have
prompted some authors to use only $T_\mathrm{e}$([\ion{O}{3}]) to calculate all
the ionic abundances. If we did this, the oxygen abundances would be higher by
up to 0.03 dex in our PNe, and by significant amounts, $0.07\mbox{--}0.20$~dex,
in the \ion{H}{2} regions: $12+\log(\mathrm{O}/\mathrm{H})'_\mathrm{CELs}=8.55$,
8.73, 8.61, 8.68, and 8.64 for M8, M16, M17, M20, and M42, respectively.
However, we consider that the evidence for the values of
$T_\mathrm{e}$([\ion{N}{2}]) shown in Table~\ref{tab1} is strong, in particular
for the \ion{H}{2} regions. Temperature values similar to those presented here
and also satisfying $T_\mathrm{e}$([\ion{N}{2}])~$>T_\mathrm{e}$([\ion{O}{3}])
are generally found at different positions within the bright areas of our sample
\ion{H}{2} regions \citep[e.g.,][]{rod99,mes08}. These temperature gradients are
also predicted by photoionization models \citep{sta78}. Furthermore, the
temperatures and temperature ratios measured for the sample \ion{H}{2} regions
from the same spectra we are using here have been shown to be consistent with
the predictions of photoionization models \citep{rod10}. We conclude that the
available evidence indicates that our values of $T_\mathrm{e}$([\ion{N}{2}]) are
reliable, but problems with this temperature could be a possible source of bias
when comparing abundances in \ion{H}{2} regions and PNe.

The presence of unresolved high-density regions in the observed volumes might also
introduce a bias in our results by affecting in different ways the CELs we are using
\citep{rub89}. Recently, \citet{tsa11} have argued that small high-density clumps
are seriously affecting the intensities of CELs in \ion{H}{2} regions, thus producing
the discrepancy between CELs and RLs in these nebulae. However, both the fact that
different density diagnostics imply very similar densities in the sample \ion{H}{2}
regions \citep{est04,gar06,gar07} and the success of constant-density
photoionization models in reproducing the intensities of the main CELs involved in
the determination of electron temperatures and oxygen abundances in these objects
\citep{rod10} suggest that high-density clumps, if present, have only a small effect
on the spectra of our sample \ion{H}{2} regions.

If we now compare our results with those derived by other authors from the same
spectra (see \citealt{tsa04} and \citealt{liu04b}, in addition to the references
listed in Table~\ref{tab1}), the differences are larger, since besides using
different atomic data, they use other lines. The differences are all
$\la0.1$~dex with three exceptions: \citet{sha04} find (O/H)$_\mathrm{CELs}$
0.21 dex below our value in IC~418, whereas \citet{tsa03} and \citet{wes04} find
(O/H)$_\mathrm{RLs}$ 0.17 and 0.21~dex above our derived values for NGC~3132 and
NGC~6543, respectively.

Finally, we note the importance of using spectra of relatively high spectral
resolution in order to obtain the best estimates for the oxygen abundance, even
when using strong CELs. If the spectral resolution is poor,
[\ion{O}{3}]~$\lambda4363$ can be blended with several lines, like
[\ion{Fe}{2}]~$\lambda4359$, \ion{O}{2}~$\lambda4367$, and
\ion{O}{1}~$\lambda4368$. In the sample \ion{H}{2} regions, these blends would
lead to values of $T_\mathrm{e}$([\ion{O}{3}]) up to 1000~K higher and final
oxygen abundances up to 0.05~dex lower. In our sample PNe the effects
are smaller: up to 300~K higher $T_\mathrm{e}$([\ion{O}{3}]) and oxygen
abundances up to 0.02~dex lower.

\section{Discussion}

\begin{figure}
\plotone{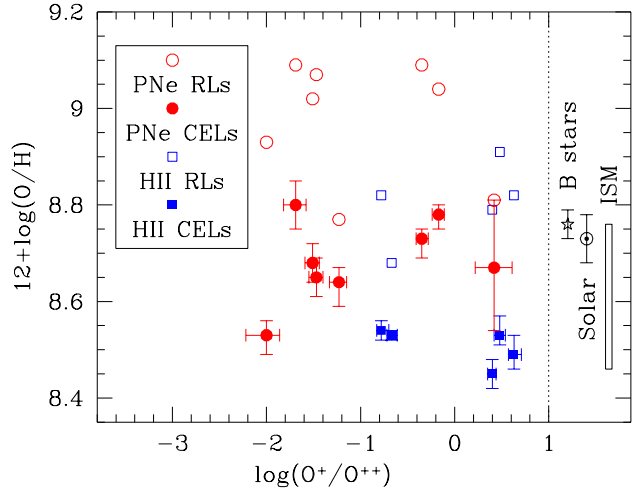}
\caption{Oxygen abundances derived using CELs (filled circles and squares) and
RLs (empty circles and squares) for our sample \ion{H}{2} regions (squares) and
PNe (circles) as a function of the values of $\mbox{O}^+/\mbox{O}^{++}$. From
left to right, the objects are: NGC~3242, NGC~6543, NGC~6210, NGC~6884,
NGC~6572, M17, M42, NGC~6720, NGC~3132, M8, IC~418, M16, and M20. We also show
the protosolar abundance \citep[the symbol $\sun$,][]{asp09},
the abundance of nearby B stars \citep[star,][]{prz08}, and the range of
abundances found for the ISM \citep[rectangle,][]{jen09}.\label{fig1}}
\end{figure}

Figure~\ref{fig1} shows the total oxygen abundances implied by CELs and RLs for
the \ion{H}{2} regions and PNe as a function of
$\mbox{O}^+/\mbox{O}^{++}$. All the \ion{H}{2} regions show similar abundances,
with
$12+\langle\log(\mathrm{O}/\mathrm{H})_\mathrm{CELs}\rangle_\mathrm{H~II}=8.52$
and
$12+\langle\log(\mathrm{O}/\mathrm{H})_\mathrm{RLs}\rangle_\mathrm{H~II}=8.80$,
suggesting that the ISM in the solar neighborhood is homogeneous. The PNe
results show more dispersion, but can be seen to be higher by about 0.2~dex:
$12+\langle\log(\mathrm{O}/\mathrm{H})_\mathrm{CELs}\rangle_\mathrm{PNe}=8.70$
and
$12+\langle\log(\mathrm{O}/\mathrm{H})_\mathrm{RLs}\rangle_\mathrm{PNe}=8.98$.
We expect to find the real oxygen abundances of the nebulae somewhere in the
ranges defined by CELs and RLs, with the different explanations of the
discrepancy favoring values at different positions along these ranges. There are
some indications that the explanations might differ in \ion{H}{2} regions and
PNe, like the fact that all the \ion{H}{2} regions studied so far show moderate
discrepancies, whereas PNe can show huge discrepancies, like NGC~1501, where
\citet{erc04} find $12+\log(\mathrm{O}/\mathrm{H})_\mathrm{CELs}=8.52$ and
$12+\log(\mathrm{O}/\mathrm{H})_\mathrm{RLs}=10.09$. However, at least three of
the PNe show CELs abundances similar to the RLs abundances of the \ion{H}{2}
regions. This implies that even if the explanation of the abundance discrepancy
is very different in each kind of object, these PNe will still show similar or
larger oxygen abundances than the \ion{H}{2} regions. This is contrary to our
expectations from simple models of galactic chemical evolution. The difference
could arise from extensive stellar migration from the inner parts of the Galaxy,
or gas flows or infall \citep[e.g.,][]{sch09}. An overabundance of oxygen in PNe
could also arise from oxygen production in the stellar progenitors, although
this is not expected to be significant at near-solar metallicities
\citep[e.g.,][]{mar01,kar10}.

In order to shed light on this issue, Figure~\ref{fig1} also shows the
protosolar abundance of \citet{asp09}, the results obtained for several nearby B
stars by \citet{prz08}, and the range of values found for the diffuse ISM by
\citet{jen09}. We will assume that these results, which can be considered as
representative of the best estimates for these objects, are not
affected by important systematic errors and we will try to find a way to
reconcile our results with them.

The RLs abundances of \ion{H}{2} regions are similar to the B-star values, but
their discrepancy with the abundances implied by CELs might be difficult to
justify, especially if one must look for a mechanism that operates in \ion{H}{2}
regions but not in PNe. One possibility is suggested by the different
sensitivity of the calculated values of (O/H)$_\mathrm{CELs}$ in \ion{H}{2}
regions and PNe to changes in the assumed temperature structure (see
Section~\ref{an}): temperature fluctuations in a chemically homogeneous medium
could be responsible for the discrepancy in \ion{H}{2} regions (supporting the
RLs abundances in these nebulae, see, e.g., \citealt{est04}), whereas some kind
of metal-rich inclusions could produce the discrepancy in PNe \citep[supporting
the CELs abundances in these objects, e.g.,][]{liu00,hen10}. However, all
these explanations have problems, related to the origin and survival of the
metal-rich inclusions or to the nature of the mechanism responsible for the
temperature fluctuations \citep[and references therein]{liu00,hen10,rod10}. We
explore below an alternative explanation.

The abundances implied by CELs in \ion{H}{2} regions are too
low, except when compared with the lower values found in the ISM. The oxygen
abundances derived by \citet{jen09} for the ISM follow a trend with the overall
amount of depletion implied by all observed elements, suggesting that the
different values represent different amounts of depletion onto dust grains.
Could the low abundances in \ion{H}{2} regions be due to their having a larger
fraction of oxygen deposited in dust grains?

Dust grains in PNe and \ion{H}{2} regions should have different characteristics,
consisting of recently created stardust in PNe, and heavily processed
interstellar grains in \ion{H}{2} regions. However, the amount of oxygen that
can be in grains outside of molecular clouds is limited by the availability of
atoms of those elements that bind to oxygen to form refractory compounds like
oxides and silicates. \citet{whi10} estimates that a maximum of 90--140 ppm of
oxygen can be located in silicates and oxides. A correction for this amount
would increase the value of
$12+\langle\log(\mathrm{O}/\mathrm{H})_\mathrm{CELs}\rangle_\mathrm{H~II}$ to a
maximum of 8.67, close to
$12+\langle\log(\mathrm{O}/\mathrm{H})_\mathrm{CELs}\rangle_\mathrm{PNe}$, but
at least some of the PNe (NGC~6210 and NGC~6543) might need similar corrections,
since they harbor oxygen-rich dust like the \ion{H}{2} regions
(\citealt{ber05}; Delgado-Inglada \& Rodr{\'i}guez, in preparation).

In fact, \citet{jen09} finds that the highest levels of oxygen depletion (the
lowest abundance values) cannot be explained with depletion in silicates and
oxides, and concludes that oxygen must be locked up with an element as abundant
as hydrogen or carbon. \citet{whi10} points out that a similar shortfall of
oxygen (around 160 ppm) is observed in molecular clouds, and argues that the
most plausible explanation involves an organic refractory dust component arising
from the irradiation of ices by UV photons in molecular clouds. This component
would not be expected to be present in PNe. So, is the missing oxygen in
\ion{H}{2} regions deposited in cometary-like dust grains?

Correcting the oxygen abundances implied by CELs in \ion{H}{2} regions by 160 ppm
(for the oxygen in the organic refractory component) and 115 ppm (as an estimate
for the oxygen in silicates and oxides), we get
$12+\langle\log(\mathrm{O}/\mathrm{H})_\mathrm{CELs}\rangle_\mathrm{H~II}=8.78$,
in agreement with the B-stars abundances. This value also agrees with the values
implied by CELs in PNe, even if they are corrected for the presence of silicates
and oxides. The latter agreement is what one would expect from the almost flat
age-metallicity relation followed by nearby F and G stars
\citep[e.g.,][]{hol09}, though the relatively small dispersion in the abundances
derived for our PNe suggests that the spread of stellar metallicities at a given
age should be smaller than what is usually considered.

In order to confirm our interpretation or to distinguish between other
possibilities, homogeneous comparisons of the abundances of other elements would
be valuable. Note, however, that they would require studying the bias introduced
by the ICFs and whether it is different in \ion{H}{2} regions and PNe.

\section{Conclusions}

We have selected a sample of five \ion{H}{2} regions and eight low-ionization
PNe that have available spectra of high quality, all of them located at
distances lower than 2 kpc. A homogeneous analysis of their oxygen abundances
based on CELs and RLs shows that the PNe are, on average, overabundant by
0.18~dex.

If we take at face value the results implied by B-stars, the Sun, and the
diffuse ISM, along with the almost flat age-metallicity relation implied by F
and G stars, we find that for the PNe, the abundances implied by CELs agree with
the expected values, whereas the abundances implied by RLs are too high. For the
\ion{H}{2} regions, the abundances implied by CELs are similar to the lower
values found by \citet{jen09} in the ISM, which are explained by \citet{whi10}
as due to depletion in organic refractory dust grains. If we assume that these
grains are also present in \ion{H}{2} regions, their CELs abundances agree with
all the other results. We can thus explain the overabundance of oxygen in PNe
through the presence of different dust components in \ion{H}{2} regions and PNe.

\acknowledgments
We thank an anonymous referee for useful comments that helped to improve this
manuscript. We acknowledge support from Mexican CONACYT projects 50359-F and
CB-2009-01/131610.

\end{document}